# A Simulation-Based Conceptual Model for Tokenized Recycling: Integrating Blockchain, Market Dynamics, and Behavioral Economics

Atta ul Mustafa[1]

**Abstract**

This study develops a conceptual simulation model for a tokenized recycling incentive system that integrates blockchain infrastructure, market-driven pricing, behavioral economics, and carbon credit mechanisms. The model aims to address the limitations of traditional recycling systems, which often rely on static government subsidies and fail to generate sustained public participation. By introducing dynamic token values linked to real-world supply and demand conditions, as well as incorporating non-monetary behavioral drivers (e.g., social norms, reputational incentives), the framework creates a dual-incentive structure that can adapt over time. The model uses Monte Carlo simulations to estimate outcomes under a range of scenarios involving operational costs, carbon pricing, token volatility, and behavioral adoption rates. Due to the absence of real-world implementations of such integrated blockchain-based recycling systems, the paper remains theoretical and simulation-based. It is intended as a prototype framework for future policy experimentation and pilot projects. The model provides insights for policymakers, urban planners, and technology developers aiming to explore decentralized and market-responsive solutions to sustainable waste management. Future work should focus on validating the model through field trials or behavioral experiments.

**Introduction**

In the circular economy, effective waste management is the key driver to achieve global sustainability aim. Despite several strategies to increase recycling through financial benefits, government subsidies, and penalties, these conventional methods failed to generate cohesive and long-term public behavior. (Dai et al., 2015; Sidique et al., 2010). With ever-increasing global waste generation, a new innovative model with a combination of financial rewards, behavioral incentives, and emerging technologies is vital to decreasing environmental pollution and increasing recycling rates (Chen & Su, 2023; Song et al., 2015).

Blockchain and tokenization is an innovative technology with the potential to improve transparency, management, and incentivization in waste collection and management system (Casino et al., 2019). Blockchain offers a transparent solution where every transaction related to recycling and collection of waste is stored in a decentralized ledger, reducing the chances of potential fraud (Reyna et al., 2018; Zheng et al., 2017). In addition, the token-based reward system ensures adaptability, linking incentives with market conditions such as demand and supply for carbon credits and recyclable materials, ensuring the receiver remains motivated to recycle at the prime times (Narayan & Brem, 2021).

[1] Doctoral Student in the School of Islamic Studies at Hamad Bin Khalifa University, Email: Atul88769@hbku.edu.qa

Although the application of token-based systems has been extensively studied in different fields, such as decentralized finance (DeFi) and supply chain management, however, token-based application in waste collection and recycling is still unexplored (Singh et al., 2020). Plastic Bank (2020) It is one of the few key studies that used token-based systems to demonstrate the potential application of blockchain technology for waste collection in emerging countries. However, a proper mathematical model is still missing in the literature for the potential of blockchain systems to be used in waste collection.

In the meantime, according to behavioral economics, recycling behavior can be influenced by non-monetary benefits as well, such as social acceptance, environmental responsibility, and social norms (Schultz, 2011; Thaler & Sunstein, 2008). Nudge theory, in particular, suggests how a slight change in the framing of incentives can make individuals do environmentally responsible actions (Allcott & Mullainathan, 2010). However, the real issue is amalgamating these behavioral patterns into a market-based tokenized system that alters incentives simultaneously in both social incentives and the market (Viscusi et al., 2013).

Furthermore, adding carbon credit to blockchain-based recycling solutions presents a better value approach to connecting recycling incentives with broader economic goals. Carbon pricing and Cap-and-Trade systems have been extensively used to mitigate greenhouse gas emissions by incentivizing carbon-neutral activities (Stavins, 2019). Incorporating carbon credit in a token-based system provides dual benefits in the form of recycling and carbon reduction, which helps attract individuals and big corporations to meet sustainability targets (Chen & Su, 2023).

This study attempts to fill the gaps in the literature by developing a cohesive tokenized recycling model with the inclusion of blockchain technology, behavioral incentives, and dynamic market pricing. The paper addresses three main questions. 1) How does tokenized pricing driven by market forces impact the participation of individuals in recycling programs? 2) What is the optimum balance between financial incentives and social incentives in driving toward recycling behavior? 3) How can carbon credit be introduced in a tokenized system to increase environmental and economic outcomes?

The remainder of this paper is organized as follows: Section 2 reviews the theoretical frameworks and relevant literature. Section 3 presents the mathematical model underlying the tokenized recycling system, including its integration with market dynamics and carbon credits. Section 4 reports the results of the Monte Carlo simulation used to analyze system performance under various market conditions. Section 5 concludes with policy implications and recommendations for future research.

**Literature Review**

The existing literature on recycling incentives primarily focuses on recycling rewards and regulatory penalties. Money incentives, in the form of deposit refunds or government subsidies, have been impactful in increasing the recycling participation rates of individuals (Dai et al., 2015; Sidique et al., 2010). The financial incentives are primarily implemented to internalize external costs to align the individual interest with the public interest, a phenomenon initially for environmental taxation by (Pigou, 1920). The studies showed that providing financial benefits for recycling does increase the recycling rate. However, after some time, it diminishes as well due to budget constraints and the static nature of the reward (Viscusi et al., 2013). Furthermore, conventional recycling systems often face strict resistance from the masses due to monetary penalties (Hage et al., 2008). These results dictate that although financial incentives increase the recycling rate, they are not sustainable because of their static nature and they do not consider individual behavioral factors and market variability.

In the recent decade, the role of behavioral economics in driving sustainability and environmentally friendly behavior has increased manifold. According to Nudge theory, a small change in social cues and default setting can significantly induce the agent to do environmental friendly behavior (Thaler & Sunstein, 2008). For example, Schultz (2011) showed that providing social feedback about the recycling works of other individuals could increase their participation rates. Similarly Muralidharan and Sheehan (2018) found that when the public perceives recycling as a socially desirable behavior, their chances of indulging in this behavior increase many times, even in the absence of good monetary rewards.

Yet, most of the research in this domain is focused on non-market scenarios primarily revolving around government-led campaigns or community-based initiatives (Hobson, 2003). These approaches highlight the potential of financial incentives and normative behavior; however, they do not integrate a market-driven system that could enhance participation through both financial and non-financial rewards.

The use of blockchain technology and tokenization has gained much attention in recent literature due to its ability to be transparent, traceable, and efficient. Blockchain technology can help to reduce uncertainty from collection to processing to recycling stage by providing traceability, while tokenization provides a new avenue for individuals and corporations to get rewards for their participation in recycling programs (Reyna et al., 2018). Unlike the conventional rewards system, which is both time-consuming and cost-intensive, the blockchain system is decentralized, and by using smart contracts, the rewards would automatically be distributed according to predefined conditions (Zheng et al., 2017).

The current financial incentives program on recycling offers fixed rewards which do not explain the market fluctuations in the demand for recyclable materials (Kinnaman et al., 2000). However, implementing a dynamic tokenization model helps to do more adaptive incentive structure which is based on the supply and demand of recycling markets (Greene & W., 2018). It could help align the individual agent behavior with the overall goal of environmental behavior and ensure that rewards remain attractive according to market conditions (Park & Zhong, 2021). The dynamic token model offers flexibility over the static incentive model by changing the rewards according to the demand of the market. When the demand for recyclable materials is high, the potential value of the token could increase to increase the financial incentives associated with recycling. Similarly, when the demand for recyclable material is low, the token value could decrease in the same fashion (Park & Zhong, 2021).

In addition to financial incentive and tokenization, the addition of carbon credit to the recycling incentive could prove to be more helpful in increasing recycling. The carbon price system and cap and trade are extensively used to reduce the carbon footprint in the ecosystem (Stavins, 2019). In addition, the recycling of materials could prove to be helpful in the reduction of carbon emissions especially in carbon-intensive industries such as plastics and metals (Song et al., 2015). Amalgamation of carbon credits into blockchain based recycling system could give two potential benefits, firstly the companies or the individuals would receive tokens for recycling, secondly, they would earn carbon credits due to the reduction in greenhouse gas emissions (Singh et al., 2020). The trade ability aspect of the carbon credit can increase the economic aspect of recycling and could attract big corporations (Chen & Su, 2023).

The above discussion highlights the need for further investigation into blockchain-based tokenization and the dynamic market-based pricing model for the recycling solution. The existing studies examined the financial incentives and behavior nudges. However, there remains an extensive gap in understanding how decentralized systems and market-based tokens could be used to solve the problem of recycling material. By addressing this gap in the literature, the study appeals to the broader aspects of environmental economics.

## Methodology

This study aims to evaluate the efficiency of a tokenized recycling incentive system by adding dynamic token pricing, behavioral economics, and carbon credits. The methodology used blockchain technology to integrate both the financial and non-financial benefits to the individuals participating in recycling work. The study used mathematical modeling, Monte Carlo simulations, and sensitivity analysis to assess the model's performance in the best and the worst scenarios.

The model used a tokenized system of rewards with dynamic market conditions, stochastic processes, and government interventions in the form of subsidies.

$$R(t) = P(t) * \eta(t) * W(t) \qquad (1)$$

R(t) is the recycling volume, which shows the total number of recycled volumes of waste at time t. P(t) is the participation rate of individuals in the recycling activity, η(t) is the efficiency of the recycling activity, and W(t) is the amount of total waste generation. So, the total amount of recycled waste depends on the participation of individuals multiplied by their efficiency and the total amount of waste.

$$\eta(t) = \eta_0 + \eta_g \cdot t$$

$\eta_0$ shows the initial efficiency of recycling activity at time 0, and $\eta_g$ is the efficiency increase due to technological advancements. The efficiency of recycling activity tends to increase over time due to technological advancements or individuals becoming experts in it.

$$dW(t) = vW(t)dt + \xi W(t)dZ_t$$

The growth of the waste generated is represented by a stochastic process to better depict real-world situations. $v$ is the growth of waste generation due to either population increase or economic growth. $\xi$ is the change in the waste generation process due to the change in consumption patterns and the economic downturn. $dZ_t$ highlights the wiener process represented by fluctuations in waste production.

The participation rate (behavioral economics) is highlighted by not only financial rewards but also because of social factors. For this purpose, the participation rate of the individuals is modeled by using the utility function to add both the financial incentives and the social motivations for the individuals to participate in this activity.

$$P(t) = P_\infty(1 - e^{-\lambda t}) * U(t)$$

$P_\infty$ is the maximum amount of participation rate or the saturation level the system can achieve. $\lambda$ shows how fast the individuals adopt the system or adoption constant. U(t) is the utility function highlighting monetary and non-monetary rewards.

$$U(t) = \alpha_1 \cdot T_v(t) + \alpha_2 \cdot S(t)$$

$T_v$ depicts the financial reward the individuals received by participating in the recycling activity. $\alpha_1$ is the relative importance of financial rewards in motivating individuals to participate in recycling activities. S(t) shows the social attention/reputation gained from participating in recycling activities. And $\alpha_2$ is the relative importance of social reputation in that regard.

Token value $T_v(t)$ (dynamic market behavior) fluctuates according to the market conditions, such as the demand and supply of recyclable markets.

$$T_v(t) = \frac{D(t)}{S(t)}$$

D(t) represents the demand for recyclable materials or the demand for tokens as a medium of exchange. While S(t) represents the supply of the token primarily induced by the issuance rate, the policy of blockchain platforms or token burns. This allows the addition of real-world market dynamics to the system as supply decreases and demand increases due to some circumstances that will cause the price of the token to rise.

However, everything comes with some cost in the form of fixed costs to run the operations and the variable costs associated with the recycled materials. This cost could be reduced by the introduction of subsidies from governments (government intervention).

$$C_o(t) = C_b + c_t * R(t) - subsidy(t)$$

$C_o(t)$ is the total operational cost at time t, $C_b$ is the initial expenses or cost needed to kickstart this system, $c_t$ is the variable cost associated with the amount of recycled materials, and subsidy (t) is the government's subsidy to decrease the burden and promote a circular economy.

Furthermore, this system also offers environmental benefits in the form of a reduction in carbon emissions due to increased recycling. Furthermore, the model adds carbon credit to gain benefits from reduction in carbon emissions.

$$E_b(t) = \alpha * R(t) + T_c * C_r(t)$$

$\alpha$ is the environmental benefit coefficient representing the environmental benefits in the form of carbon emission reduction per unit of recycled materials. $T_c$ is the carbon credit providing the financial benefits for carbon reduction and $C_r(t)$ is the recycled material that qualifies for the carbon credit.

After defining the parameters now, we can look at the total recycling volume and the token revenue

$$R(t) = P_\infty(1 - e^{-\lambda t}) * U(t) * (\eta_0 + \eta_g.t) * (W_0 + \gamma t)$$

So, the total amount of recycled materials depends on the maximum participation rate of individuals, utility function, and the amount of total waste generated, increasing stochastically over time. For the token revenue

$$R_t(t) = R(t) * T_v(t)$$

Where $T_v(t)$ is the dynamic token value that changes according to supply and demand, and R(t) is the volume of recycled materials.

For the net benefit, we combine token revenue, environmental benefits, and operational costs.

$$N_b(t) = P_\infty(1 - e^{-\lambda t}) * U(t) * (\eta_0 + \eta_g.t) * (W_0 + \gamma t) * \left(\frac{D(t)}{S(t)} + \alpha + T_c - c_t\right) - C_b + Subsidy(t)$$

The total net benefits increased with an increase in participation rate, token value, and environmental benefits while decreasing with an increase in operational cost.

To get the optimal participation rate, we need to differentiate this equation with respect to P(t) and solve it by putting it 0

$$\frac{N_b(t)}{\partial P(t)} = (1 - e^{-\lambda t}) * U(t) * (\eta_0 + \eta_g.t) * (W_0 + \gamma t) * \left(\frac{D(t)}{S(t)} + \alpha + T_c - c_t\right) = 0$$

$$\frac{D(t)}{S(t)} + \alpha + T_c - c_t = 0$$

$$T_v(t) = c_t - \alpha - T_c$$

So, it suggests that for an optimal participation rate, the token value must be sufficient to cover the operational costs, environmental benefits, and carbon credits. After holding this condition true, the participation rate will peak.

Furthermore, we also conducted a sensitivity analysis to examine how changes in the key parameters (Carbon tax, governmental subsidies, token value, and operational cost) may change the net benefit. This helps us to understand which variable could impact the most and which have potential risk capabilities. Firstly, with token value.

$$\frac{\partial N_b(t)}{\partial T_v(t)} = P(t) * (\eta_0 + \eta_g.t) * (W_0 + \gamma t)$$

According to the equation, increasing the token value increases the net benefit from the system linearly; a higher token value makes the system look more attractive to the individuals making them participate more and could generate higher income. For operational costs, we have

$$\frac{\partial N_b(t)}{\partial c_t} = -P(t) * (\eta_0 + \eta_g.t) * (W_0 + \gamma t)$$

The equation dictates that the net benefit received from the system will decrease with increasing operational costs. So, it is better to go for more innovative technology to reduce operational costs or offset them with governmental subsidies. Similarly, for carbon credit and carbon pricing, we have

$$\frac{\partial N_b(t)}{\partial T_c} = P(t) * (\eta_0 + \eta_g.t) * (W_0 + \gamma t)$$

The impact of the increase in carbon credit is just like an increase in token revenue and will result in an increase in net benefit because it provides more incentives to recycle materials to earn financial rewards. For government subsidies, we have

$$\frac{\partial N_b(t)}{\partial Subsidy(t)} = 1$$

As we can see from the equation, the increase in government subsidies will directly and positively impact net benefit by reducing operational costs.

The next step in the model is the introduction of Monte Carlo simulation. The study used 10,000 simulations where the primary variables, such as participation rate, token value, waste generation, and carbon credit, are taken as random variables from a proper probability distribution. There are some assumptions as well. The token value is a lognormal distribution depicting the real market volatility. The participation rate used beta distribution to anticipate the adoption uncertainty; operational cost follows a normal distribution, representing the fluctuations in the operational cost parameter, and carbon credit follows a normal distribution, highlighting the introduction of future regulatory requirements. Running Monte Carlo simulations with different scenarios helps us to understand how well the model will perform in different conditions and uncertainties.

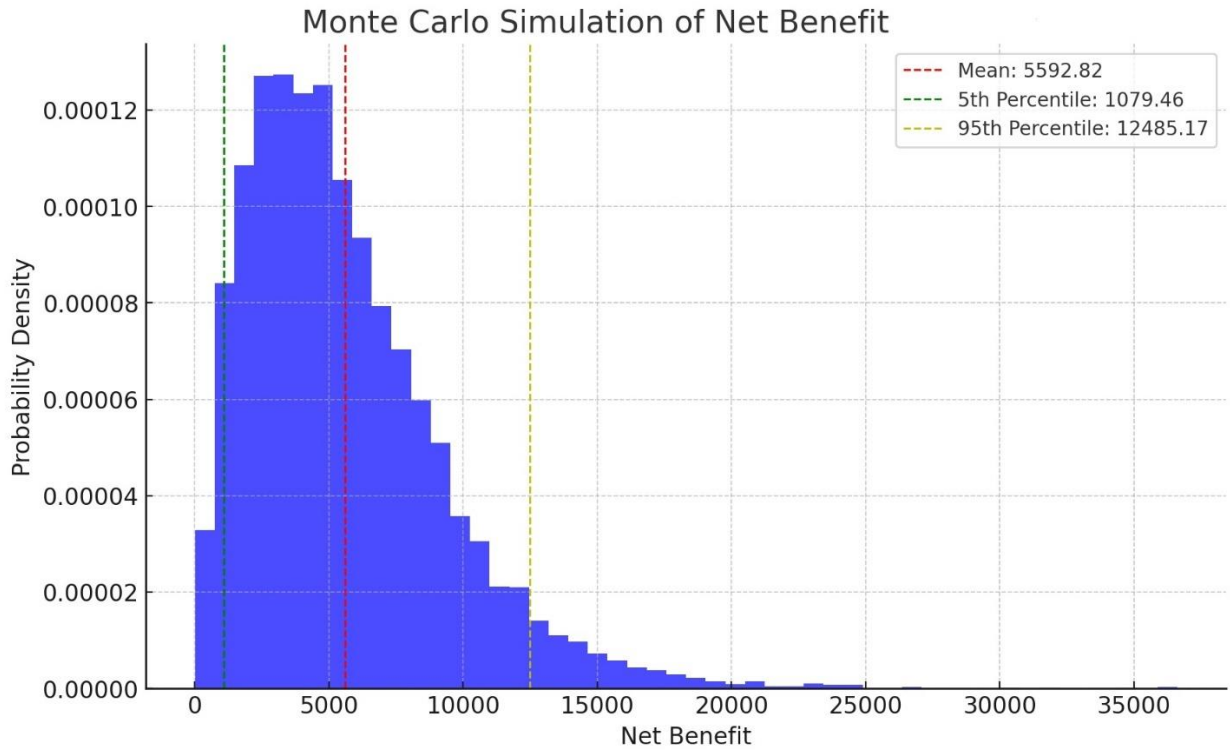

The above graph depicts the simulation analysis using the Monte Carlo simulation. As we can see from the graph, the system generates overall positive impacts throughout the different scenarios. The 5th percentile shows pessimism, while the 95th percentile shows optimism. The lowest value is 1079.49, and the highest is 12485.1, with a mean of 5592.82 and a standard deviation of 3702.48. The standard deviation value highlights the many fluctuations in the model primarily driven by the market fluctuations of token value or the participation rate. However, even in the worst scenario, the positive overall economic benefit makes it sustainable against unfavorable conditions.

**Tokenized Recycling system versus Subsidy based system**

In this comparative simulation analysis of a tokenized recycling incentive model versus a traditional subsidy-based model, key assumptions were established to closely approximate real-world dynamics of recycling systems, focusing on operational costs, participation rates, and incentives.

**Common Assumptions Across Models**

- **Operational Cost**: Both models commence with a baseline operational cost of 50,000 units, which represents the core financial requirements for recycling operations. To simulate market fluctuations, this cost is assumed to vary by a 2% standard deviation, accounting for potential cost changes associated with labor, materials, and other operational expenses.
- **Participation Rate**: An average participation rate of 50% was assumed across both models, representing moderate public engagement in recycling activities. This serves as a neutral benchmark for comparing how model-specific incentives impact recycling behavior.
- **Carbon Credit Incentive**: In line with contemporary environmental objectives, the tokenized model uniquely includes a carbon credit incentive, assigning a value of 5 units per recycled item. This incentive reflects additional financial rewards for environmental contributions (e.g., emissions reduction). The subsidy model does not incorporate carbon credits, maintaining a traditional reward structure that lacks environmental incentives.
- **Standard Recycling Volume**: For comparative consistency, both models assume a uniform recycling volume of 1,000 units. This standardization permits an equivalent analysis of cost and revenue effects across models, regardless of scale.

**Tokenized Model-Specific Assumptions**

- **Market-Driven Token Value**: The tokenized model operates with a dynamic token value, initially set at 15 units, which fluctuates according to a log-normal distribution (mean multiplier of 2). This variability models real-world market dynamics by aligning incentive value with fluctuations in recyclable material demand and supply, thus increasing or decreasing the financial appeal of recycling.
- **Participation Rate Adaptation**: Unlike the static subsidy model, the participation rate in the tokenized model is positively correlated with token value changes. As token value rises above its base level, public engagement correspondingly increases, reflecting a market-responsive incentive structure that encourages higher recycling rates when financial returns are favorable.

**Subsidy Model-Specific Assumptions**

- **Fixed Financial Reward**: The subsidy-based model provides a constant financial reward of 10 units per recycled item, simulating a traditional incentive system that remains unaffected by market conditions or demand fluctuations. This static reward structure reflects the typical rigidity of subsidy programs.
- **Static Participation Rate**: The participation rate in the subsidy model remains at the baseline 50%, as the model does not incorporate dynamic or market-responsive incentives.

This static engagement level is consistent with the subsidy model's limited reliance on financial variability to motivate recycling behavior.

The simulation analysis revealed notable differences in economic outcomes, participation rates, and environmental impact between the tokenized and subsidy-based recycling models. The tokenized model demonstrated a mean net benefit of 67,501 units, indicating substantial economic viability despite fluctuating market conditions. In contrast, the subsidy model resulted in a mean net benefit of -45,007 units, underscoring the financial limitations of a static reward system that lacks adaptability to market shifts.

The participation rate in the tokenized model proved to be dynamic, increasing in response to higher token values, which suggests that public engagement can be enhanced by aligning incentives with market-driven fluctuations. This responsiveness to market conditions contrasts sharply with the subsidy model, where participation remained constant at the baseline rate of 50%, as the fixed financial rewards did not stimulate additional engagement.

Furthermore, the environmental impact of the tokenized model was strengthened by the inclusion of carbon credits, which provided an additional financial reward for each recycled unit and aligned individual recycling efforts with broader sustainability goals. The subsidy model, devoid of such environmental incentives, did not contribute additional benefits beyond the fixed monetary reward for recycling. Overall, the tokenized model's flexibility, economic viability, and alignment with environmental objectives suggest that it may offer a more sustainable and adaptive approach to incentivizing recycling than traditional subsidy-based models.

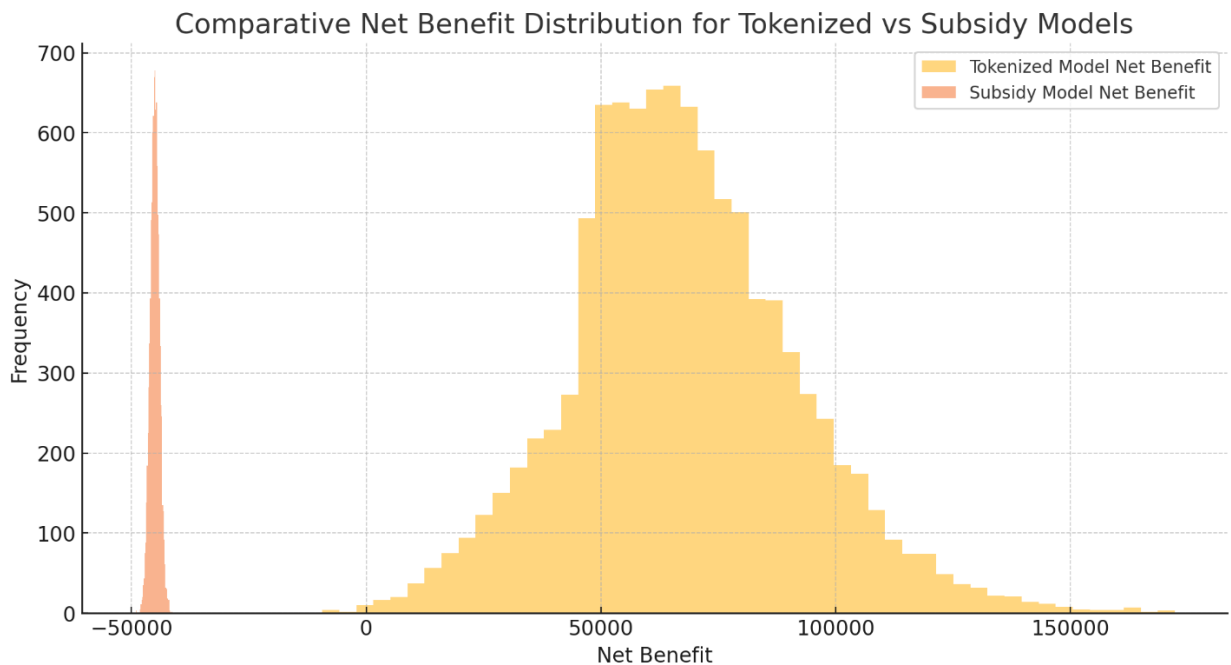

*Figure 1: Tokenized versus Subsidy Model*

**Conclusion**

The study develops a robust mathematical model for recycling material by integrating blockchain technology, behavioral economics, and real-world market prices. The model tries to address the gaps in the conventional recycling models, which are mainly based on government subsidies, rewards, and penalties, and they fail to generate long-term public participation in recycling activities. The traditional methods suffer due to budget constraints, and the public loses interest in these models due to static rewards and insufficient monetary gains.

This paper addresses these limitations by providing a model based on blockchain technology backed by a decentralized ledge system to increase model efficiency and transparency. In addition, the amalgamation of tokens as a financial reward adds a dynamic system of price mechanisms determined by the supply and demand for recycling materials. This dynamic system ensures that the system remains attractive and the participation rate remains stable throughout the process.

In addition, the model incorporates carbon credit, linking the recycling materials with the broader environmental goals to make it relevant for global sustainability goals. By giving dual benefits, the models try to combine individual micro agents and corporation macro agents into a single model. Introducing carbon credit is especially beneficial for large corporations because they are concerned about meeting environmental goals, and carbon credit is the way to go. In this way, both individuals and corporations stand together for larger environmental impacts.

According to the model results, it remains economically beneficial with positive values even in the worst scenario. However, the model exhibits a large standard deviation, suggesting many fluctuations in the price of token value, which could make predicting the participation of individuals in the recycling activity a bit difficult. However, the overall positive value suggests that individuals will remain willing to participate in recycling.

The implementation of blockchain-based infrastructure for waste management is essential. Governments and private entities should invest in these platforms to establish transparent, traceable, and efficient systems for waste collection and recycling, enhancing trust and accountability by tracking every stage of the recycling process. Additionally, transitioning from fixed monetary rewards to a dynamic token-based incentive system will help maintain the attractiveness of recycling incentives, ensuring sustained participation from the public and businesses. The value of these tokens should be aligned with the demand and supply conditions of recyclable materials to optimize economic benefits.

**Policy implications**

Implementing a blockchain-based token system for recycling management is very beneficial, and it could help the government, and the concerned authorities track the process through every stage of the recycling process, increase its efficiency and transparency, and increase people's trust. Furthermore, it will help the government waste management process to remain competitive and attractive to the general people.

Furthermore, it is beneficial for the government to combine the token-based recycling system with the carbon credit policy. The world is continuously getting more and more attention regarding the emissions of large corporations in their manufacturing process. This setup will help the government make it more beneficial for large corporations to participate in the recycling process. In the same way, it will help the country to achieve environmental goals.

The government needs to start collaborating with technology providers and large corporations to make it sustainable and effective. Furthermore, it is imperative for the government to kickstart the process as the initial cost is always high, and the government could do it in the form of subsidies, tax rebates, or policy support to make the system more attractive for adoption.

The government may need to start giving the individuals working in the recycling materials process a special citizen status. It will help to generate a behavioral response, and the remaining population will start to get attracted to these titles. Furthermore, integrating financial rewards with non-financial rewards will create a more suitable scenario for individuals to participate more in this process.

**Limitations and Future Research**

While the proposed model presents a novel integration of blockchain, tokenization, behavioral economics, and carbon credits for waste management, it remains entirely conceptual and simulation-based. No real-world data currently exists for such decentralized token-based recycling schemes, and the model's parameters are derived from analogous domains (carbon markets, token economies, and behavioral interventions).

Future research should focus on (i) piloting small-scale implementations of blockchain-based recycling rewards, (ii) collecting empirical data on public adoption and token valuation behaviors, and (iii) refining the simulation model using real market and behavioral response data. Collaboration between municipalities, blockchain developers, and behavioral researchers will be critical for real-world validation and policy adoption.